\documentclass[journal=nanolett,manuscript=article]
{achemso}
\usepackage{graphicx}
\usepackage{dcolumn}
\usepackage{bm}
\usepackage{amsmath}
\usepackage{amssymb}
\usepackage{hyperref} 
\DeclareGraphicsExtensions{.pdf,.eps,.png,.jpg,.mps}
\usepackage{mathrsfs}
\usepackage{soul} 
\usepackage[utf8]{inputenc} 
\usepackage{hyperref} 
\usepackage{xcolor} 
\usepackage{eucal}
\usepackage{setspace}
\usepackage{pifont}
\definecolor{rojo}{RGB}{255, 0, 0}

\definecolor{azul}{RGB}{0, 0, 255}

\definecolor{carrotorange}{rgb}{0.93, 0.57, 0.13}

\usepackage{flushend}

\usepackage[version=3]{mhchem}

\usepackage{booktabs}

\title[]{Thermal Hofstadter Butterflies}

\abbreviations{DOS}

\author{Natalia Cortés} 
\email{natalia.cortesm@usm.cl}
\affiliation{Departamento de Física, Universidad Técnica Federico Santa María, 2390123 Valparaíso, Chile}

\author{Bastian Castorene}
\affiliation{Instituto de Física, Pontificia Universidad Católica de Valparaíso, Casilla 4950, 2373223 Valparaíso, Chile}
\altaffiliation{Departamento de Física, Universidad Técnica Federico Santa María, 2390123 Valparaíso, Chile}

\author{Francisco J. Peña} 
\affiliation{Departamento de Física, Universidad Técnica Federico Santa María, 2390123 Valparaíso, Chile}

\author{Damian Melo} 
\affiliation{Instituto de Física, Pontificia Universidad Católica de Valparaíso, Casilla 4950, 2373223 Valparaíso, Chile}
\altaffiliation{Departamento de Física, Universidad Técnica Federico Santa María, 2390123 Valparaíso, Chile}

\author{Sergio E. Ulloa} 
\email{ulloa@ohio.edu}
 \affiliation{Department of Physics and Astronomy and Nanoscale and Quantum Phenomena Institute, Ohio University, Athens, Ohio 45701, USA}
 
 \author{Patricio Vargas}
\affiliation{Departamento de Física, Universidad Técnica Federico Santa María, 2390123 Valparaíso, Chile}

\begin{document}

\begin{abstract}

Fractal electronic spectra arising from the competition between lattice periodicity and magnetic flux are a fundamental hallmark of two-dimensional quantum systems. While the spectral properties of Hofstadter butterflies are well documented, their thermodynamic response has remained remarkably unexplored. We present an original characterization of the electronic entropy $S_{e}$, and specific heat $C_{e}$, at half-filling, for square, honeycomb, and triangular lattices under a magnetic field. We demonstrate that these observables exhibit fast and slow magneto-thermo oscillations and pronounced magnetocaloric effects.
We identify striking self-similarity in $S_e$ and $C_e$, tracing heart-shaped specific heat and tunnel-like entropy contours that repeat at specific lattice-dependent magnetic fluxes.  
Entropy minima at low temperatures play a remarkable role, acting as fingerprints for the butterfly spines, resolving the underlying fractal spectra. These findings may establish thermal measurements as high-resolution spectroscopic probes, providing a robust framework for recognizing fractal signatures through thermodynamics in diverse nanostructures.

\end{abstract} 
 
\maketitle

\section{Introduction}

In 1976, Douglas Hofstadter published his seminal work exploring the spectral properties of non-interacting electrons confined to a two-dimensional (2D) square lattice subjected to a uniform magnetic field \cite{hofstadter1976}. This model has since become a cornerstone study in a wide variety of systems, with lattice periodicities ranging from  nanometers in quantum materials to centimeters in classical setups. This special spectral system has motivated explorations in a host of modern condensed matter systems, resulting in new insights on quantum Hall phases, moiré heterostructures, ultracold fermion gases, and topological materials. 
The model has inspired a wide array of research, from Chern number classifications and quantum Hall physics~\cite{thouless1982quantized}, to recent 
observations in moiré superlattices \cite{andrei2021,dean2013,ponomarenko2013,Hunt2013,Nuckolls2025}, and 
in engineered cold-atom systems in optical lattices~\cite{aidelsburger2013,miyake2013,PhysRevA.89.011603}, microwave cavities \cite{Schuster2018,Ritsch2019}, two-dimensional photonic systems \cite{Rabl2021}, and acoustic crystals \cite{acoustic2019}.

\medskip

Hofstadter's work centers on the importance of the energy spectrum of the system depending sensitively on whether the magnetic flux per lattice period $\alpha$ is a rational or irrational number. For rational values of $\alpha$, the system obeys Bloch's theorem and exhibits a band structure with an integer number of magnetic subbands, such that the whole spectrum exhibits a beautiful fractal structure, famously known as the \textit{Hofstadter butterfly}. This intricate spectrum, which can be obtained by numerically solving for the energy eigenvalues as a function of rational values of $\alpha$, illustrates the rich self-similar pattern that encodes the interplay between lattice periodicity and magnetic translation symmetry.
The Hofstadter butterfly exemplifies how concepts from number theory (e.g., Diophantine equations and continued fractions) naturally emerge in the spectral analysis of quantum systems with appropriately commensurate parameters. 

Despite extensive studies of its spectral properties and topological classification, the thermodynamic response of electrons of the Hofstadter butterfly's fractal energy landscape has remained largely unexplored. Early work calculated magnetization and specific heat at low temperatures in the square lattice, revealing the characteristic gap structure through oscillations in these observables \cite{Xiang2008,yang2012specificheat}. How lattice geometry shapes thermodynamic response, however, has not been systematically addressed.

This work provides the first theoretical characterization of Hofstadter butterflies under temperature in square, honeycomb, and triangular geometries, bridging the gap between abstract fractal energy spectra and measurable thermodynamic quantities, and contributing with fertile ground for exploring further physical observables in these 2D lattice geometries.

Through the use of single-orbital tight-binding equations and Fermi-Dirac statistics, we perform a detailed analysis of the thermal behavior for electrons as a function of magnetic flux $\alpha$ and temperature $T$ for the three lattice geometries across different parameter regimes. We present results for the electronic specific heat $C_e$, and electronic entropy $S_e$. 
We find that the thermodynamic response reflects universal thermal effects for the three lattices around flux $\alpha=1/2$, as well as unique features at particular $\alpha$ numbers for each lattice geometry. Both thermal quantities reveal fractal signatures in each spectrum through intricate patterns arising from the coupling between $\alpha$ and $T$, highlighting phenomena such as fast and slow magneto-thermo oscillations and large magnetocaloric effects. 
These results may provide a clear roadmap for the experimental verification of the fractal nature of the spectra.

\begin{figure*}[t]
\includegraphics[width=\linewidth]{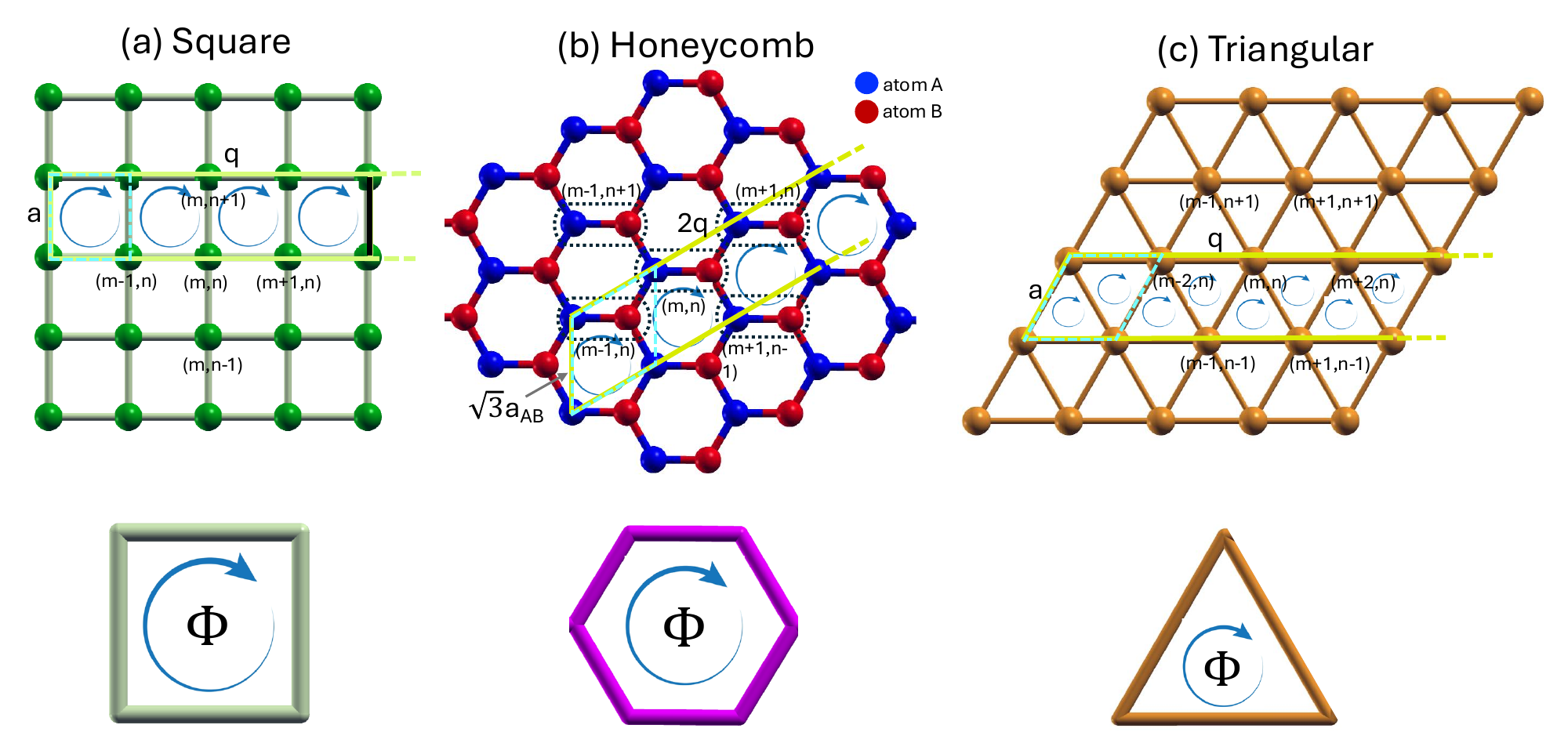}
\caption{Top panels: (a) square, (b) honeycomb, and (c) triangular geometrical lattices. The light green rectangle in (a) and parallelograms in (b) and (c) show the magnetic unit cells, the cyan square in (a) and parallelograms in (b) and (c) represent the geometrical unit cells, $a_{\text{AB}}$ is the distance between A-B atoms in b. Ordered pairs label the nearest neighbor sites to the $(m,n)$ site in each lattice. Bottom panels show the corresponding elemental plaquettes with magnetic flux $\Phi=\alpha \Phi_0$, with rational $\alpha=p/q$.}\label{fig1} 
\end{figure*}

\section{Energy spectra} \label{sec:spectra}

The allowed eigenstates of an electron in a uniform magnetic field in a given lattice geometry are determined through single-band tight-binding models to capture the fine structure of the spectrum and describe its recursive properties. The single-band tight-binding Hamiltonian $H$ considers a uniform magnetic field $\mathbf{B} = B \hat{z}$  employing the Peierls substitution \cite{Peierls1933}. In this framework, the electronic hopping matrix elements acquire a phase factor due to the associated vector potential $\mathbf{A}$, leading to the (spinless) Hamiltonian

\begin{equation}\label{Hamiltonian}
H = -t \sum_{\langle i, j \rangle} e^{i \theta_{ij}} c_i^\dagger c_j,
\end{equation}
where the sum runs over all nearest-neighbor pairs $\langle i,j \rangle$ in the lattice, and $c^{\dagger}_{i}$ and $c_{j}$ correspond to fermionic operators creating and annihilating an electron at site $i$ and $j$, respectively. The hopping amplitude $t$ between nearest neighbors will be used as the unit of energy throughout the paper. The Peierls phase $\theta_{ij}$ is given by
\begin{equation}\label{Peierlsphase}
\theta_{ij} = \frac{2\pi}{\Phi_0} \int_{\mathbf{r}_i}^{\mathbf{r}_j} \mathbf{A} \cdot d\mathbf{r},
\end{equation}
with $\Phi_0 = h/e$ the magnetic flux quantum, $h$ the Planck constant, and $e$ the elementary charge. For each geometry in Fig.\ \ref{fig1}, $d\mathbf{r}$ parametrizes the path between two lattice sites, $\mathbf{r}_i$ and $\mathbf{r}_j$.

As we explicitly show in the supplementary document \cite{SI}, solving the Schrödinger equation $H\psi= E \psi$, with $H$ in Eq.\ \ref{Hamiltonian}, and $\varepsilon=E/t$, requires defining a relevant rational dimensionless parameter $\alpha$ as \cite{hofstadter1976}
\begin{equation}\label{Eq:alpha}
\alpha = \frac{\Phi}{\Phi_0}=\frac{p}{q},
\end{equation}
where $\Phi$ is the magnetic flux through a single lattice plaquette (see Fig.\ \ref{fig1}), and 
$p$ and $q$ are relative prime integers. This parameter provides essential details in the problem: 
it enumerates the magnetic subbands in each system, as $q$ in the square and triangular lattices (and $2q$ in the honeycomb case); it also allows characterizing the symmetry of the energy spectrum under inversion about  $\alpha=\frac{1}{2}$ for each lattice system.
Because of this symmetry, it is convenient to divide the three lattice geometries in Fig.\ \ref{fig1} into two groups: symmetric and anti-symmetric `butterflies', based on the reflection symmetries of their energy spectrum with respect to the $\alpha=\frac{1}{2}$ value.  

Numerical diagonalization of the matrices associated with the eigenvalue equation for each lattice yields the eigenenergies $\{\varepsilon_l(\alpha)\}$, where the index $l$ labels the states for each $\alpha$ value.
These calculations result in the well-known symmetric spectra for the square and honeycomb lattices, whereas the antisymmetric butterfly spectrum is that of the triangular lattice.  The supplement contains the tight-binding equations that yield the different lattice spectra needed to compute the thermodynamic response for each butterfly \cite{SI}. 
   
\section{Butterfly Thermodynamics}

The three geometries described above and in the supplement exhibit different thermodynamic properties. 
We obtain the electronic internal energy $U_e$ (not shown), specific heat $C_e$, and entropy $S_e$, for different magnetic flux per plaquette $\alpha$, and temperature $T$. For concreteness, we consider the half-filling condition $\nu=\frac{1}{2}$ for both symmetric and nonsymmetric butterflies, although other filling factors can be easily studied.

To calculate thermodynamic quantities, it is convenient to define the density of states (DOS) by
\begin{equation}\label{eq:dos}
D_\alpha(\varepsilon) = \sum_{l=1}^{Q} \delta\big(\varepsilon - \varepsilon_l(\alpha)\big),
\end{equation}
where $l$ runs from 1 to $Q$ for each value of $\alpha$, with $Q=q$ for the square and triangular lattice, and $Q=2q$ for the honeycomb lattice.  
A crucial characteristic of the square and honeycomb lattice spectra is that the reflection symmetry about $\varepsilon=0$ ensures perfect electron-hole symmetry in the DOS, $D_\alpha(\varepsilon) = D_\alpha(-\varepsilon)$. On the contrary, $D_\alpha(\varepsilon) \neq D_\alpha(-\varepsilon)$ for the triangular lattice (see supplement \cite{SI}).
We should note, however, that the full spectrum is traceless, so that $\sum_{l=1}^Q \varepsilon_l=0$ for each $\alpha$ value in all three lattices.
As we show below, these symmetry properties of the DOS are useful for setting the chemical potential $\mu$ in the systems.

\subsection{Chemical Potential}

In a half-filled system, $\nu=\frac{1}{2}$, only the lower half of the $Q$ energy levels are occupied in the ground state of the symmetric energy spectra of the square and honeycomb lattices. 
Using the DOS in Eq.\ \ref{eq:dos}, the chemical potential at finite $T$ satisfies the condition
\begin{equation}\label{eq:filling_condition}
\langle N \rangle=\int 
D_{\alpha}(\varepsilon)f(\varepsilon,\mu,T)d\varepsilon=\sum_{l=1}^{Q} f(\varepsilon_l(\alpha), \mu, T) = \nu Q,
\end{equation}
where $f(\varepsilon,\mu,T)=[1+e^{(\varepsilon-\mu)/k_B T}]^{-1}$  is the Fermi-Dirac distribution function. 
The chemical potential $\mu = 0$ satisfies the half-filling condition, Eq.\ \ref{eq:filling_condition}, for all values of $\alpha$ and $T$ due to the electron-hole symmetry of the spectra in the square and honeycomb lattices. 

\begin{figure}
\centering
\includegraphics[width=\columnwidth]{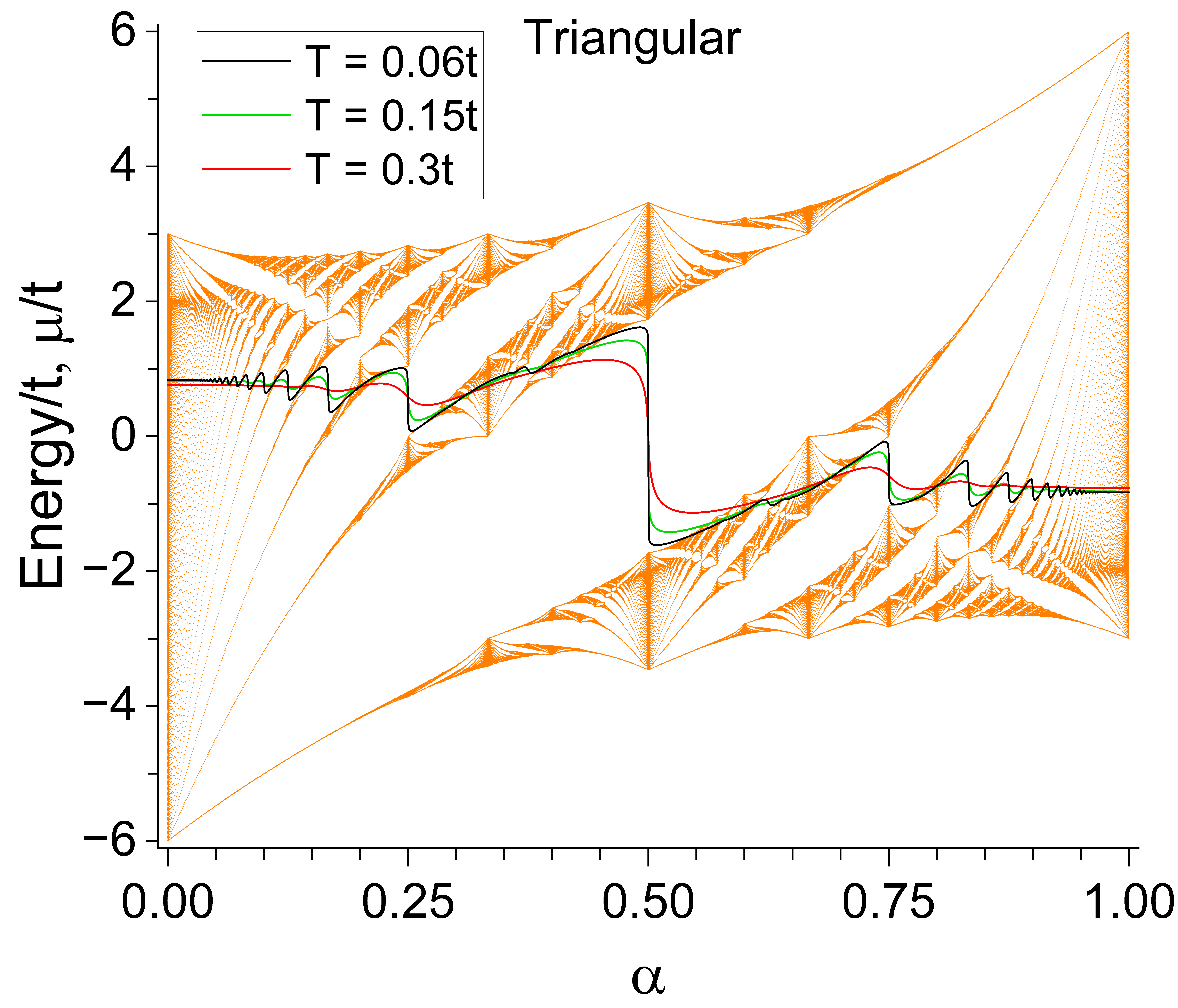}
\caption{Chemical potential curves $\mu(\alpha,T)$ (solid lines) as a function of $\alpha=p/q$ ($q=1123$) and different values of temperature $T$ under the condition of half-filling in the triangular lattice; all energies in units of the hopping parameter $t$.  The energy spectrum $\varepsilon$ is shown behind the curves, where the sharp drops in $\mu$ are seen to occur within the gaps in the spectrum.}\label{fig:chempot_triangular} 
\end{figure}

The corresponding occupation constraint that defines the chemical potential $\mu =\mu(\alpha, T)$ in the triangular lattice is given also by Eq.\ \ref{eq:filling_condition}.
Due to the spectral asymmetry of the triangular lattice butterfly, $\mu$ is seen to change its value with magnetic flux $\alpha$, even at zero temperature, to satisfy the half-filling condition.  This is exemplified by the $\mu$ curves in Fig.\ \ref{fig:chempot_triangular}, which show sharp changes whenever the spectrum has gaps.
It is interesting to observe that $\mu(\alpha,T)$ in Fig.\ \ref{fig:chempot_triangular} is antisymmetric with respect to $\alpha=1/2$, so that $\mu(\frac{1}{2}-\delta,T)=-\mu(\frac{1}{2}+\delta,T)$, with $\delta$ rational, and that this relation is valid for all temperatures. \footnote{This condition is connected to the antisymmetry of the spectrum, $-\varepsilon_l(\frac{1}{2}+\delta)=\varepsilon_{q-l}(\frac{1}{2}-\delta)$, which together with its null trace, $\sum_l \varepsilon_l=0\; \; \forall \alpha$, results in symmetric thermodynamic functions, such as $C_e(\frac{1}{2}+\delta)=C_e(\frac{1}{2}-\delta)$, as seen in the results shown.}
The strong variation of $\mu$ with $\alpha$ reflects the non-monotonic DOS of the spectrum---a characteristic also present in the symmetric butterflies, but hidden for the case of half-filling there ($\mu = 0$). 

\subsection{Thermodynamic Quantities}
The chemical potential $\mu=0$ for symmetric butterflies, and $\mu(\alpha,T)$ for the antisymmetric butterfly, all at half-filling, are used to calculate the electronic specific heat $C_e$ and entropy $S_e$ shown below. The electronic specific heat is calculated through its elementary definition $C_e=\partial U_e/\partial T$ 
\begin{equation}\label{eq:internal_energy}
C_e(\alpha,T)=\frac{\partial}{\partial T} \int 
\varepsilon D_\alpha(\varepsilon)f(\varepsilon_, \mu, T)d\varepsilon,
\end{equation}
and the entropy in terms of the Fermi-Dirac function $f$ is
\begin{equation}\label{entropy}
S_{e}(\alpha,T)= -\int 
D_\alpha(\varepsilon)[f\ln f+(1-f)\ln(1-f)]d\varepsilon.
\end{equation}
The DOS in Eqs.\ \ref{eq:internal_energy} and \ref{entropy} is given by Eq.\ \ref{eq:dos}. The supplement contains detailed information for analytical and numerical thermodynamic quantities for all butterflies \cite{SI}.

\section{Finite-temperature results}

The thermodynamic calculations are performed for all rational flux values of $\alpha$  $\in [0,1]$ (Eq.\ \ref{Eq:alpha}), with $q=1123$ and $p\ \text{integer} \in [0,q]$. 
Choosing a large prime number $q$ ensures both computational tractability and sufficient resolution to capture the fractal structure of the three lattice butterflies, together with their thermodynamic response at all temperatures. 
Note that as discussed by Hofstadter,\cite{hofstadter1976} large values of $q$ provide a well-bounded approximation of the spectrum even for irrational values of $\alpha$.  This is reflected in the smooth behavior of the various thermodynamic quantities when shown as a function of $\alpha$, as we will see below. \footnote{To capture the rich $T$ dependence of thermodynamic functions, we employ an exponential mesh, $T_k=T_1(T_2/T_1)^{(k-1)/(N_p-1)}$, with $k=1,...,N_p$.  Typically, $N_p=200$, $T_1=0.01t$ and $T_2=1t$.}

\subsection{$T$-$\alpha$ thermodynamic contours}
We display specific heat and entropy contour maps over the $T$ and $\alpha$ plane for each of the three lattice geometries.  These representations allow one to correlate the spectral characteristics, such as gaps and high DOS regions, with clear features of the thermodynamic functions, especially at low temperatures.  

\paragraph{Square lattice.}
\begin{figure*}[tb]
\centering
\includegraphics[width=\textwidth]{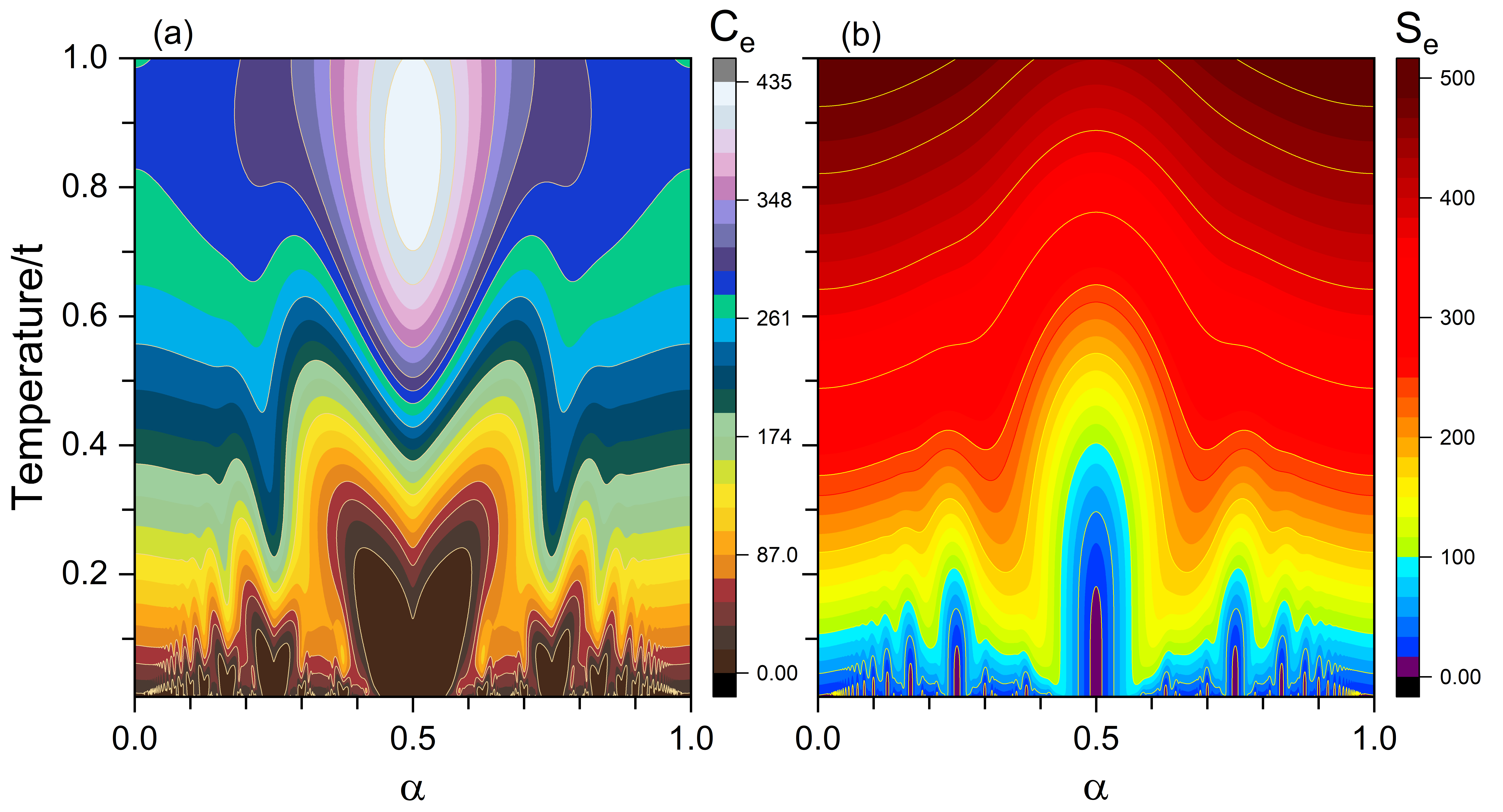}
\caption{{\bf Square lattice}. Contour plots for (a) electronic specific heat $C_e(\alpha, T)$, and (b) entropy $S(\alpha, T)$,  as function of the magnetic flux parameter $\alpha=p/q$ ($q=1123$), and temperature $T/t$. Solid light-yellow lines represent selected constant-$C_e$ or constant-$S_e$ contours. Color bars in units of $k_B$. Notice rich oscillatory structure associated with spectrum.}\label{fig4} 
\end{figure*}

Figure \ref{fig4} presents results for the square lattice system, the original butterfly.  Both thermodynamic functions, $C_e$ and $S_e$ are symmetric about $\alpha=\frac{1}{2}$, inheriting the symmetry of the energy spectrum (Fig.\ S1a in supplement \cite{SI}).  At low temperatures ($T \lesssim 0.2t$) these functions are most sensitive to the gap structure near $\varepsilon=0$, exhibiting intricate fast magneto-thermo oscillations coincident with the spectral features.  Constant specific heat curves (some indicated by light-yellow lines) in panel \ref{fig4}a appear as heart-like shapes of different sizes, depending on the corresponding nearby spectral gaps; the largest `heart' centered at $\alpha=\frac{1}{2}$. As temperature increases, thermal broadening smooths out fine-scale structures, leading to smaller isothermal variations and a more regular behavior for $C_e$ contours. Notice that the sharp low-$T$ oscillations are accompanied by deep broad minima in $C_e$ (dark brown) that evolve into maxima at higher $T$ values, especially clear near $\alpha=\frac{1}{2}$.

The entropy for the square lattice in \ref{fig4}b shows deep minima with `tunnel'-like shapes near spectral gaps (purple zones), centered at the butterfly `spines' (for values $\alpha = \frac{1}{2}, \frac{1}{4},...$).  The repeating features in the isentropic curves as $T$ increases (indicated by yellow lines) in this case appear as blunt, long peaks extending into the blue zones, with sizes that correspond to the nearby spectral gaps as well. Unlike the specific heat, constant-temperature entropy curves show sharp minima at the butterfly spines, especially clear at low $T$.  Although progressively smoother at higher temperature, the constant-$T$ entropy curves continue to show minima in those regions up to $T\approx 0.6t$, but the minima around $\alpha=1/2$ are preserved across all $T$. The entropy is expectedly larger at higher temperatures, it shows less oscillatory structure, with local maxima around $\alpha=0,1$, and becoming nearly featureless for $T\gtrsim 1t$.

\paragraph{Honeycomb lattice.}

\begin{figure*}[tb]
\includegraphics[width=\textwidth]{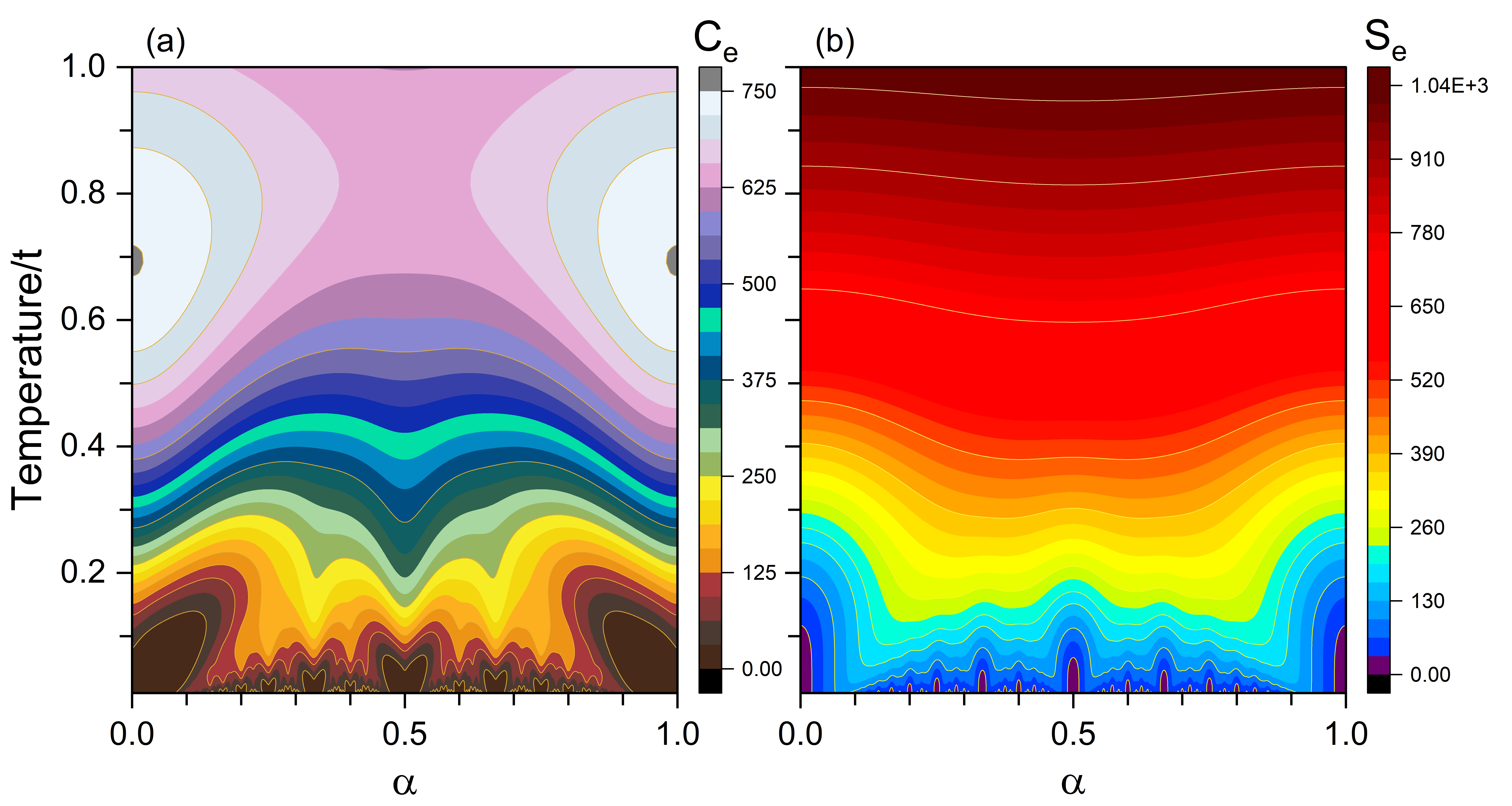}
\caption{{\bf Honeycomb lattice}. Contour plots for (a) electronic specific heat $C_e(\alpha,T)$, and (b) entropy $S_e(\alpha, T)$ as function of the magnetic flux parameter $\alpha=p/q$ ($q=1123$), and temperature $T/t$. Solid light-yellow lines represent selected constant-$C_e$ or constant-$S_e$ contours. Color bars in units of $k_B$.}\label{fig5} 
\end{figure*}

Figure \ref{fig5} shows results for the symmetric honeycomb butterfly (Fig.\ S1b in supplement\cite{SI}). The specific heat and entropy functions are symmetric about $\alpha=\frac{1}{2}$, as expected.  At low temperatures, they show interesting oscillations that coincide with the intricate spectral structures.  Constant-$C_e$ curves in this case (light-yellow lines) have also heart-like shapes, although the associated minima are not as broad as for the square lattice, given the overall smaller gaps in the spectrum near $\varepsilon=0$, and are correspondingly displaced in $\alpha$, see Fig.\ S1b. The low-$T$ minima for $C_e$ near $\alpha=0,1$ evolve into maxima at high temperature, centered near $T=0.7t$. Notice also that the $C_e$-contours minima at $\alpha=1/2$ change to maxima near $T=0.7t$, highlighting the presence of the dense central butterfly spine, accessible at high $T$.

The isentropic contours in the honeycomb lattice in Fig.\ \ref{fig5}b, show tunnel-like shapes at low $T$ (purple zones) with clear local peaks centered also on the butterfly spines, and with sizes that indicate the overall density of the spectral honeycomb features (blue contours). As $T$ increases, close to $T=0.3t$, the isentropic curves lose their oscillatory behavior across $\alpha$, and become nearly featureless at much lower temperatures than in the square lattice.

\paragraph{Triangular lattice.}

\begin{figure*}[h]
\centering
\includegraphics[width=\textwidth]{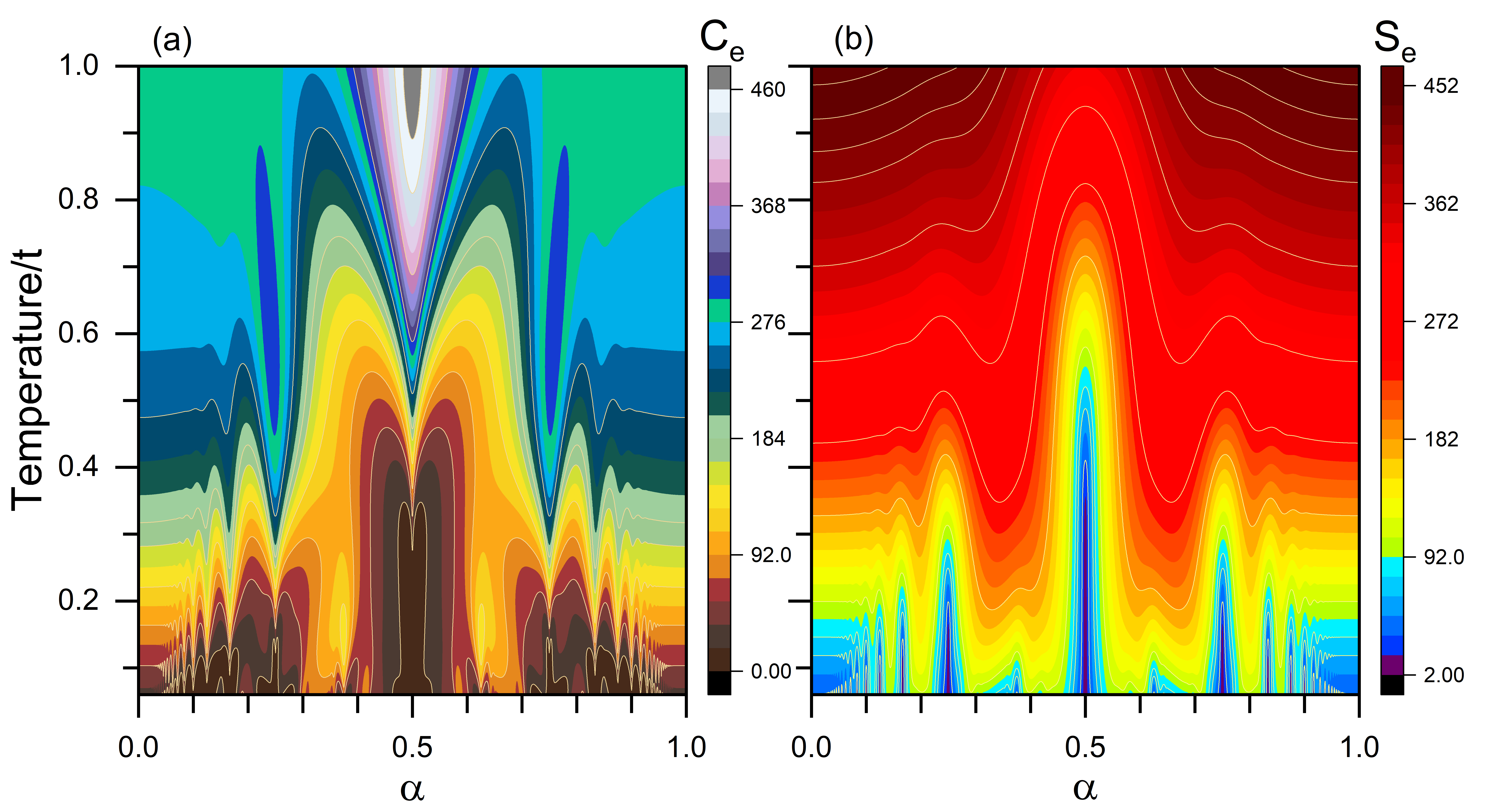}
\caption{{\bf Triangular lattice}. Contour plots for (a) electronic specific heat $C_e(\alpha, T)$, and (b) entropy $S_e(\alpha, T)$, as function of the magnetic flux parameter $\alpha=p/q$ ($q=1123$), and temperature $T/t$. Solid yellow lines represent selected contours in both panels. Color bars in units of $k_B$.}\label{fig6} 
\end{figure*}

Despite the lower symmetries in the spectrum of the triangular lattice seen in Fig.\ \ref{fig:chempot_triangular}, the thermodynamic functions such as $U_e$,  $C_e$ or $S_e$ recover inversion symmetry with respect to $\alpha=\frac{1}{2}$ at this filling factor. $^1$ 

It is interesting to see narrow and deformed heart-like shapes here, with nonsymmetric minima (except at $\alpha=\frac{1}{2}$), defined by the constant-$C_e$ curves in Fig.\ \ref{fig6}a, especially at low $T$ (shown with light-yellow lines). These minima appear whenever $\mu$ traverses a subband gap in the spectrum, for $\alpha=\frac{1}{2},\frac{1}{4},...$, see Fig.\ \ref{fig:chempot_triangular}. The deformed hearts also delimit regions of low specific heat values (dark brown), flanked by high value regions (yellow), whenever $\mu$ is in a dense subband region. As before, local minima of $C_e$ evolve into local maxima at higher temperatures, reaching a hot spot at $\alpha=\frac{1}{2}$ near $T\simeq1t$. Thermal excitations into dense spines in the triangular spectrum take over, making $C_e$ contours smooth as $T$ increases.

The entropy in the triangular lattice, Fig.\ \ref{fig6}b, shows deep minima within the 'tunnels' (purple regions) whenever $\mu$ is in a gap, that clearly accompany the pattern of hearts seen in panel a.  Similar to $C_e$, here fast oscillations at low $T$ for small and large $\alpha$ appear (light yellow lines), which become nearly flat as $T$ increases. The low-$T$ isentropes define sharp peaks on this map, centered at spectral gaps where $\mu$ drops, see Fig.\ \ref{fig:chempot_triangular}, that persist over relatively large temperatures, before becoming smoother for $T\gtrsim 1t$. 

\subsection{Entropy oscillations and magnetocaloric effect}

It is important to analyze the physical significance of the interestingly complex behavior of the isentropes in the different lattices in Figs.\ \ref{fig4}, \ref{fig5} and \ref{fig6}, panels b. The calculated electronic entropy $S_e(\alpha, T)$ provides a clear map of the system's magnetocaloric potential for cooling and/or heating as the magnetic field changes~\cite{deOliveira2010_magnetocaloric,ram2018_magnetocaloric,negrete2018_magnetocaloric_antidot,cortes2022proximity}.
By following the isentropes (light yellow lines), one can observe the temperature response of each lattice to variations in the magnetic flux $\alpha$. The resulting variations in $T$ required to maintain a constant $S_e$ indicate that each system undergoes heating and cooling cycles as the flux is modulated. The strong entropy variations as $T$ changes are associated with the clear convergence of states as flux changes, producing characteristic alternating regions of high and low DOS near $\mu$ for each lattice.
The square and triangular geometries show fast $S_e$ oscillations for extreme $\alpha$ fluxes. Notably, those steepest temperature gradients occur at low $T$, where the entropy is more strongly affected by the fine structure of the energy spectrum. The honeycomb lattice shows large $S_e$ variations near extreme $\alpha$, with `tunnel'-like isentropes increasing as $T$ rises.

The square and triangular butterflies exhibit larger magnetocaloric effect (MCE), all centered at the spines $\alpha=\frac{1}{2},\frac{1}{4},...$, where small flux changes are needed for an isentrope to be preserved, producing large variations in $T$ (blue and yellow $S_e$ gradients in the maps). Particularly, approaching $\alpha = \frac{1}{2}$, the convergence of isentropes toward temperature minima suggests a high cooling efficiency near this flux value for all lattices. The pronounced temperature variations of the isentropes signify a robust MCE, which could, in principle, have important applications in thermal engines that utilize these systems. 
As pointed out above, however, as $T$ increases, the distinct quantum signatures are masked by thermal broadening, leading to an evolution from a highly oscillatory adiabatic response to a more monotonic regime and a less efficient MCE. 

\subsection{Fractal signatures in different lattices}

\begin{figure*}
  \centering
    \includegraphics[width=1\linewidth]{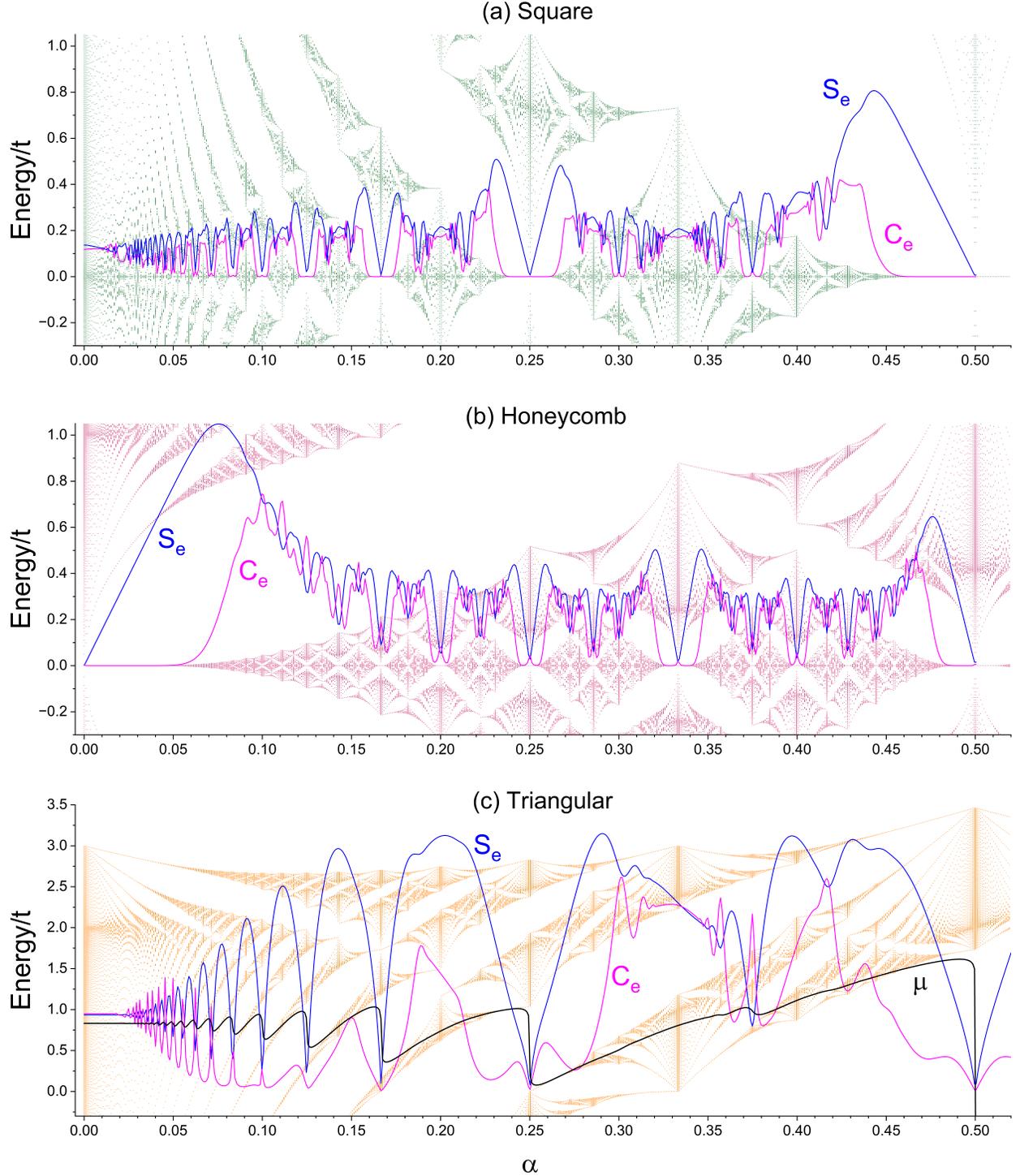}
  \caption{Electronic entropy $S_e$ (blue lines) and specific heat $C_e$ (magenta lines) isotherms superimposed on a portion of the energy spectrum. (a) Square lattice, (b) honeycomb lattice, (c) triangular lattice; all at half-filling. Isotherms in (a)-(b) are for $T=0.01t$, plots are $TC_e$ and $TS_e$; for (c) $T=0.06t$ and shown are $TC_e/2$, and $TS_e/2$ to fit scale. $k_B=1$.
 Note that minima in $S_e$ occur for $\alpha$ values associated with the `spine' of the butterflies, such as $\alpha=\frac{1}{4}, \frac{1}{2},...$. Differences seen among different lattices are closely related to their spectra near the Fermi level. 
 See text for discussion.}
  \label{fig7}
\end{figure*}

Figure \ref{fig7} displays isotherms for the electronic specific heat ($T C_e$) and entropy ($T S_e$) as a function of the magnetic flux parameter $\alpha$ at fixed low $T$ for all three lattices. These curves are superimposed
on a portion of the energy spectrum to facilitate the identification of the thermodynamic behavior with spectral fractal features of each lattice. As we consider half-filling in all cases, the chemical potential for the square and honeycomb lattices (a and b panels) is constant at $\mu(\alpha)=0$ (not shown), while for the triangular lattice in (c), $\mu(\alpha)$ is indicated by a black line (that starts nonzero at $\alpha=0$, and after oscillations reaches $\mu(\alpha=\frac{1}{2})=0$, and changes sign beyond, as discussed above).

It is interesting to notice a number of similarities and differences among the various lattice isotherms $S_e$ and $C_e$, all connected to features of the fractal spectrum given by the low temperature values used here ($T=0.01t$ in a, b; $T=0.06t$ in c). The thermal energy $k_B T$ acts as a resolution limit in all cases. Since the thermal energy is much smaller than the primary gaps ($\Delta \varepsilon_{gap}$), the major features of the butterflies are well-resolved.

The thermal response is mostly determined by the DOS near the chemical potential $\mu$. For small $\alpha$, the square and triangular lattices have nearly constant $C_e$ and $S_e$ isotherms, as the DOS at $\mu$ is nonzero. However, the honeycomb lattice shows vanishing $C_e$, but linear $S_e$ with minima located at specific $\alpha$ values, indicating the vanishing DOS near $\mu$. The pattern of vanishing specific heat and linear entropy is clearly repeated for the square and honeycomb lattices, whenever $\mu$ is near a minimal DOS or gap.
These drops occur as the low DOS negates (or suppresses) thermal excitations at low temperatures, which changes as $\alpha$ shifts to higher DOS regions in the spectrum.

The minima for $S_e$ at the butterfly `spines' represent a clear signature of fractality, which are evident for $\alpha=\frac{1}{8}, \frac{1}{6}, \frac{1}{4}, \frac{1}{2}$ in panel a; and $\alpha= \frac{1}{4}, \frac{1}{3}, \frac{1}{2}$ in panel b. These patterns for the magnetic flux in the square lattice as $\alpha=\frac{1}{2n}$, and $\alpha=\frac{1}{n}$ in the honeycomb lattice (with integer $n>0$), allow one to indirectly `observe' the underlying butterfly spectrum by looking for entropy minima. 

It is also interesting to notice oscillations in $C_e$ and $S_e$ whenever $\mu$ traverses a dense subband region in the spectrum. These can be seen in the square lattice for $\alpha\simeq0.2, 0.285, 0.33, 0.4$, and repeated at even smaller $\alpha$ values in an evidently self-similar pattern (although the oscillations at smaller level spacings for lower flux are eventually washed away by thermal broadening).  In the honeycomb lattice, similar oscillations appear in paired structures near flux $\alpha\simeq0.18, 0.22, 0.285, 0.375,...$, smoothly evolving into featureless curves for smaller $\alpha$. 

The triangular lattice is rather different, as the chemical potential traverses larger gaps as the flux is changed to maintain half-filling.  The total vanishing of the DOS midgap produces sharp minima in both $C_e$ and $S_e$, without the nearly flat plateaus seen in $C_e$ in panels a and b. For small $\alpha$, the sharp minima in $C_e$ transition to peaks that coincide with $S_e$ minima, showing complementary oscillations for both thermal quantities. This is clearly seen for $\alpha \leq \frac{1}{10}$. Notice also that panel c shows a higher $T$ isotherm (six times higher than in panels a and b), which smooths out many features. However, the self-similar oscillations that occur whenever $\mu$ traverses a dense subband region are also visible (here near $\alpha=\frac{1}{3}, \frac{1}{5}$).  As mentioned before, the lower symmetry of the triangular lattice spectrum and the varying $\mu$ at half-filling result in a more complex behavior in the thermodynamic quantities, as the butterfly patterns are compressed and stretched in non-symmetric fashion, when compared to those in the square and honeycomb lattices. 

All of these results demonstrate that even at finite, but low, temperatures, the electronic entropy and specific heat serve as sensitive probes of the Hofstadter butterfly spectral patterns. The close correspondence between thermodynamic minima and spectral gaps, as well as regions of dense subband levels, provides evidence that $S_e$ and $C_e$ encode the fractal structure of the spectrum in all of these lattices.

\section{Conclusions}

The thermodynamic responses we captured in the electronic specific heat $C_e$ and entropy $S_e$ produce insightful information for each butterfly spectrum, seen as distinctive oscillations and repeated patterns at special magnetic flux values $\alpha$ depending on the lattice geometry.
At low and moderate temperatures these patterns in $T$-$\alpha$ contour maps show `heart'-like shapes for the specific heat, and `tunnel'-like isentropic shapes of diverse sizes and widths. The largest $C_e$ hearts are symmetric and centered at $\alpha=1/2$, with deep minima inside them in all lattices. However, the symmetry of the hearts is broken for $\alpha \neq 1/2$ in the triangular lattice due to its varying chemical potential at half filling.    
The tunnels in $S_e$ contours still are symmetric as $\alpha$ varies, with isentropes converging toward temperature minima near mid flux for the three lattices, indicating a high cooling efficiency with only small flux variations. 
The honeycomb lattice in particular exhibits large isentropic changes near extreme flux values $\alpha=0,1$, suggesting its utility in adiabatic cooling cycles at accessible fluxes. 
All these results reveal a robust magnetocaloric potential inherent in these fractal electronic systems. 

The specific low-$T$ limit as $\alpha$ is varied can serve as an alternative window into the fractal nature of the spectra. Specifically, entropy minima provide a clear fingerprint for the butterfly spines at characteristic flux $\alpha = 1/2,1/4,1/6,...$ for the square lattice, and $\alpha = 1/2,1/3,1/4,...$ for the honeycomb lattice. Our calculations would help identify these repetition patterns where thermal fractal signatures emerge, and assist in comparison with experiments.

Looking ahead, the recent prediction of Hofstadter butterflies in 2D covalent organic frameworks (COFs) presents promising material systems with intrinsically large lattice constants \cite{Cuniberti-COF2023}, so that accessible magnetic fields are required to achieve high flux $\alpha$ regimes, opening an experimental scenario for the observation of Hofstadter butterflies  in 2D systems. Other large scale covalently linked organic structures (CLOS) may also allow studies at accessible magnetic fields.\cite{OQC2021}
While isolating the electronic contribution from a 2D lattice background remains an experimental challenge, yet recently achieved \cite{aamir2021ultrasensitive}, the wide availability of CLOS and COFs with large lattice constants may offer interesting possibilities to explore some of our predictions. Measurements of specific heat and entropy in these modern 2D materials may provide diagnostics of fractality, demonstrating that thermal quantities can serve as high-resolution spectroscopic probes of fractal thermodynamics.

\section{Acknowledgments}

N.C. acknowledges the financial support from Fondecyt Grant No. 11221088 during the years $2023-2025$. N.C., F.P., S.E.U., P.V., acknowledge the financial support from Fondecyt Grant No. 1240582. B.C. acknowledges the support of ANID Becas/Doctorado Nacional 21250015. D.M. acknowledges the support of ANID Becas/Doctorado Nacional 21252315.
N.C. acknowledges hospitality during a visit to Ohio University, where this research was initiated; and Amanda and Reza Alamzadeh for valuable conversations. P.V. acknowledges ANID grant CIA250002 (CEDENNA).

\clearpage

{\centering
\section{\Huge{Supplemental Information}}}

\section{Symmetric Butterflies}

The presence of magnetic field $\mathbf{B}$ breaks the discrete translational symmetry
of electrons in 2D lattices (in addition to time-reversal symmetry), due to the associated vector potential $\mathbf{A}$. One can then construct a magnetic supercell, as shown in Fig.\ 1 in the text, that recovers translational invariance for an appropriate choice of gauge (see below). 
These choices lead to magnetic phases in the Hamiltonian that produce energy spectra with reflection symmetry with respect to both zero energy and $\alpha=1/2$, as discussed before \cite{hofstadter1976,satija2025}.

\begin{figure*}
\centering
\includegraphics[width=\linewidth]{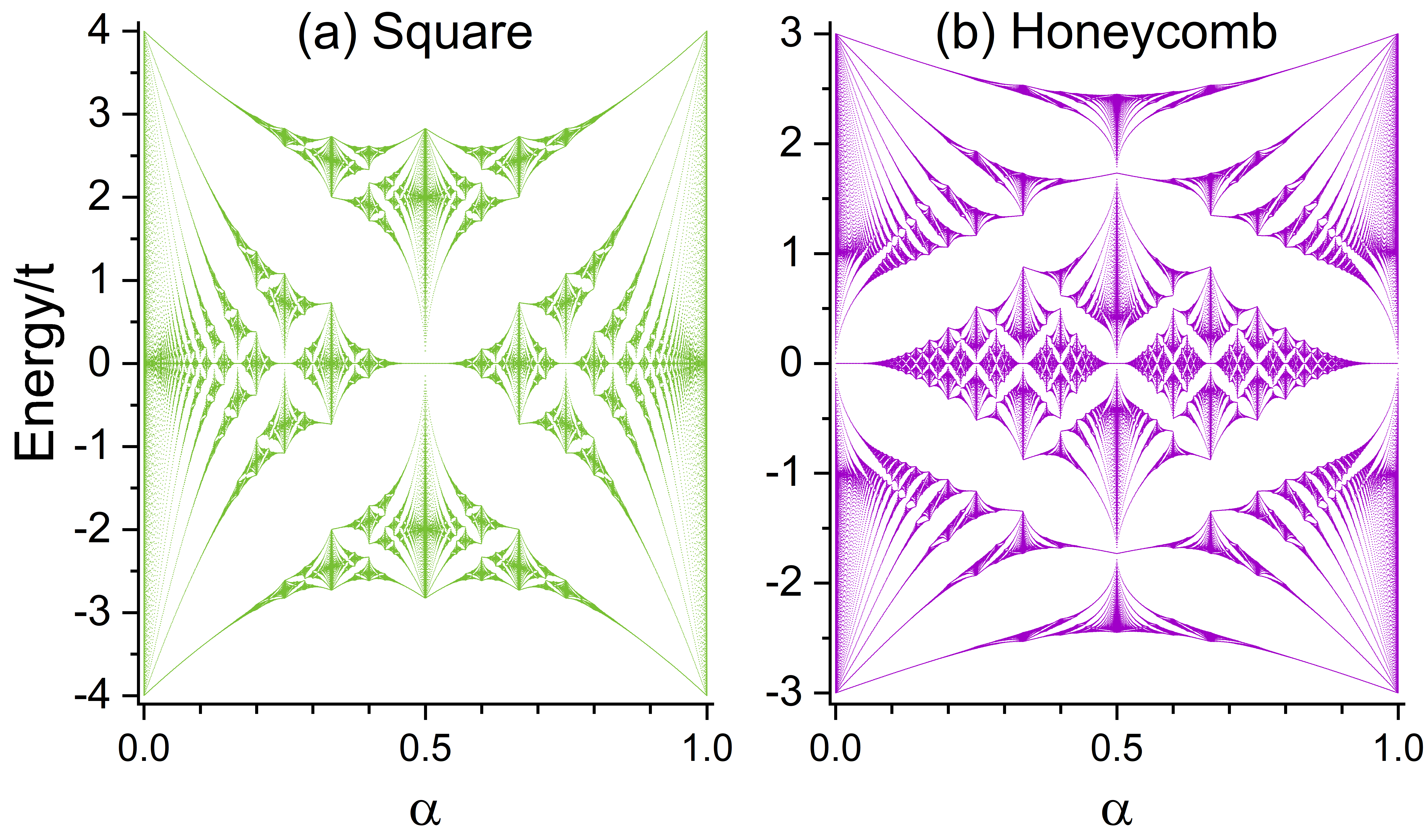}
\caption{Electronic energy spectra in units of the hopping parameter $t$ as a function of $\alpha=p/q$, for (a) square lattice, and (b) honeycomb lattice. For these spectra $\alpha=p/q$,  $p=0,1,..,q$, and $q=1123$.}\label{fig2} 
\end{figure*}

\subsection{Square lattice}
It is convenient to use the Landau gauge for the vector potential $\mathbf{A}=(0,Bx,0)=Bx\hat{y}$.  The Hamiltonian keeps translational symmetry along the $y$-direction and the phases in Eq.\ 2 in the text only appear for hops along the $y$ direction.  The wavefunction at each site $\mathbf{R}_{mn}=ma\hat{x}+na\hat{y}$ (with lattice constant $a$) is then given by $\Psi(m,n)=\psi(m)\, e^{ik_y na}$, where $k_y$ is the conserved momentum, and the tight-binding equation reads \cite{hofstadter1976}
\begin{equation}\label{Eq:square}
    \psi(m-1) + 2\cos(2\pi m \alpha-k_y)\psi(m)+\psi(m+1)=\varepsilon \psi(m),
\end{equation}
with energy eigenvalues $\varepsilon=E/t$. The magnetic flux per square plaquette in this geometry is $\Phi=BS=Ba^2$, and we have set $a=1$ in Eq.\ \ref{Eq:square}.  Notice that Eq.\ \ref{Eq:square} is invariant under the change $m \rightarrow m+q$, for rational values of $\alpha$.
This guarantees $\psi$ to fulfill a periodicity condition $\psi(m)=\psi(m+q)$, and defines a magnetic supercell of length $q$ along the $x$ direction.

\subsection{Honeycomb lattice}

The tight-binding model for the honeycomb lattice includes sublattices A and B in the hexagonal unit cell, Fig.\ 1b in the text. Using the vector potential $\mathbf{A}=Bx\hat{y}$ leads to different phases along particular hoppings in the hexagonal geometry \cite{canonico2018shubnikov}. 
Using these phases, and the conservation of $k_y$ momentum, one can decouple the resulting tight-binding equations for the two sublattices, and obtain separate relations for each of them.  For example, for sublattice B, $\Psi^B(\mathbf{R}_{mn})=e^{ik_y y_{mn}}\, \psi(m)$, one can write
\begin{multline}\label{Eq:honeycomb}
     e^{ik_y/2} u_m \psi(m-1)+u_{m+1}v_{m+1}\psi(m)
     +e^{-ik_y/2}u_{m+1}\psi(m+1)
     = (\varepsilon^2 -1)\psi(m), 
\end{multline}
with coefficients
\begin{align}
    u_m=2\cos\Big(\frac{k_y}{2}-\alpha \pi(m-\frac{1}{6})\Big),\nonumber \\
    v_m=2\cos\Big(\frac{k_y}{2}+\alpha \pi (m-\frac{1}{6})\Big).
\end{align}
The index $m$ labels the position of a single A-B dimer at $\mathbf{R}_{mn}=m\mathbf{a}_1+n \mathbf{a}_2$, with $\mathbf{a}_1=a(\sqrt{3},1)/2$, $\mathbf{a}_2=a(0,1)$, and the honeycomb lattice constant $a$ has been set to 1 in Eq.\ \ref{Eq:honeycomb}. The magnetic flux in this geometry is $\Phi=BS$, with $S=\sqrt{3}a^2/2$ the area of the hexagonal plaquette, Fig.\ 1b. 
Notice that the magnetic supercell of period $2q$ is implicit in the coefficients, with $u_m=u_{m+2q}$, and $v_m=v_{m+2q}$, and the wave functions fulfill the condition $\psi(m)=\psi(m+2q)$.      

The coefficients in Eqs.\ \ref{Eq:square} and \ref{Eq:honeycomb} are unique for each geometry, containing information about the lattice periodicity and the magnetic length scale. Each value of $\alpha$ defines a system of coupled equations and an associated matrix of size $q \times q$ (square), or $2q \times 2q$ (honeycomb). Choosing a large prime number, such as $q=1123$, yields a dense eigenvalue spectrum, as shown in Fig.\ \ref{fig2} (for $k_y=0$). These plots exhibit fascinating self-similar structure, as well as reflection symmetries with respect to zero energy and half flux quantum.

\section{Antisymmetric Butterfly}

\begin{figure}
\centering
\includegraphics[width=\columnwidth]{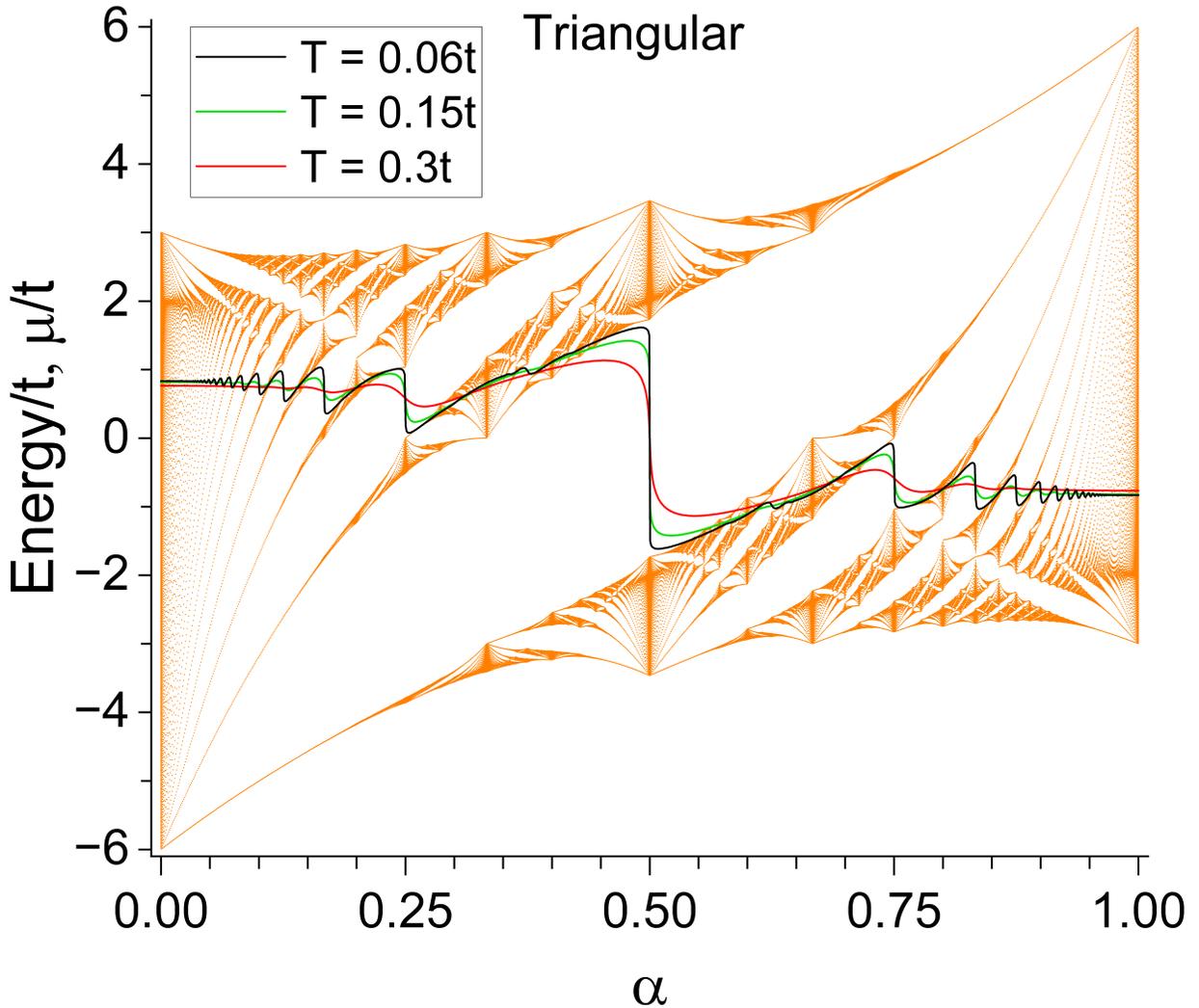}
\caption{Chemical potential curves $\mu(\alpha,T)$ (solid lines) as a function of $\alpha=p/q$ ($q=1123$) and different values of temperature $T$ under the condition of half-filling in the triangular lattice; all energies in units of the hopping parameter $t$.  The energy spectrum $\varepsilon$ is shown behind the curves, where the sharp drops in $\mu$ are seen to occur within the gaps in the spectrum.}\label{fig:chempot_triangular} 
\end{figure}

Electrons in a triangular lattice in the presence of $\mathbf{B}$ were studied by F. H. Claro and G. H. Wannier \cite{claro1979magnetic}. Similar to the square lattice system, the vector potential $\mathbf{A}=(0,Bx,0)$ generates magnetic phases along diagonal hoppings in the triangular geometry \cite{oh2001comment}. 
Using these phases and translation symmetry along $y$, the tight-binding equation for the triangular lattice is
\begin{equation}\label{Eq:triangular}
    \psi(m-2)+u_{m-1}\psi(m-1)+u_m \psi(m+1)+\psi(m+2)=\varepsilon\psi(m),
\end{equation}
with coefficients
\begin{equation}
    u_m=2\cos\Big(\frac{\sqrt{3}}{2}k_y-2 \pi\alpha(m+\frac{1}{2})\Big),
\end{equation}
The index $m$ labels sites in the magnetic supercell following the triangular geometry in Fig.\ 1c. The magnetic flux is $\Phi=BS$, with $S=\sqrt{3}a^2/4$ the area of a triangular plaquette of side $a$. The period of Eq.\ \ref{Eq:triangular} is implicit in the coefficient $u_m=u_{m+q}$, and the wave functions fulfill $\psi(m)=\psi(m+q)$.     
The matrix representation for Eq.\ \ref{Eq:triangular} has size $q \times q$, and yields a non-symmetric energy spectrum that however shows rich self-similar structure, as seen in Fig.\ \ref{fig:chempot_triangular}. We note that symmetry about $\alpha=1$ is seen when $\alpha$ runs from 0 to 2.\cite{claro1979magnetic}

\section{Density of States}

Numerical diagonalization of the matrices associated with Eqs.\ \ref{Eq:square}, \ref{Eq:honeycomb}, and \ref{Eq:triangular}, yields the eigenenergies $\{\varepsilon_l(\alpha)\}$, where the index $l$ runs from 1 to $Q$ for each value of $\alpha$, with $Q=q$ for the square and triangular lattice, and $Q=2q$ for the honeycomb lattice. The density of states (DOS) is given by
\begin{equation}\label{eq:dos}
D_\alpha(\varepsilon) = \sum_{l=1}^{Q} \delta\big(\varepsilon - \varepsilon_l(\alpha)\big).
\end{equation}
A crucial characteristic of the square and honeycomb lattice spectra is that the reflection symmetry about $\varepsilon=0$ ensures perfect electron-hole symmetry in the DOS, $D_\alpha(\varepsilon) = D_\alpha(-\varepsilon)$. On the contrary, $D_\alpha(\varepsilon) \neq D_\alpha(-\varepsilon)$ for the triangular lattice. 
We should note, however, that the full spectrum is traceless, so that $\sum_{l=1}^Q \varepsilon_l=0$ for each $\alpha$ value in all three lattices.
As we show below, these symmetry properties of the DOS are useful for setting the chemical potential $\mu$ in the systems.

\section{Thermodynamic Quantities Symmetric Butterflies}

The electronic specific heat is calculated through its elementary definition $C_e=\partial U_e/\partial T$, with $U_e$ the electronic internal energy 
\begin{equation}\label{eq:internal_energy}
U_{e}(\alpha,T)= \int 
\varepsilon D_\alpha(\varepsilon)f(\varepsilon_, \mu, T)d\varepsilon 
=\sum_{l=1}^{Q} \varepsilon_l(\alpha) f(\varepsilon_l(\alpha), \mu, T),
\end{equation}
where we have made use of Eq.\ \ref{eq:dos}. 
Using the established condition of half filling yielding $\mu=0$, the specific heat has the form
\begin{equation}\label{eq:specific_heat_general}
    C_e(\alpha, T) = \sum_{l=1}^{Q} \varepsilon_l(\alpha) \frac{\partial f_0}{\partial T} = \sum_{l=1}^{Q}\frac{\varepsilon_l^2(\alpha)}{4T^2 \cosh^2\left(\frac{\varepsilon_l(\alpha)}{2T}\right)},
\end{equation}
with $f_0 \equiv f(\varepsilon_l(\alpha), \mu=0, T)$, and $k_B=1$. 
The electronic entropy $S_e(\alpha, T)$ is given by
\begin{equation}\label{entropy}
S_{e}(\alpha,T)= -\int 
D_\alpha(\varepsilon)[f\ln f+(1-f)\ln(1-f)]d\varepsilon,
\end{equation}
in terms of the Fermi-Dirac function $f$. With the DOS in Eq.\ \ref{eq:dos}, and the condition $\mu=0$, $S_e$ can be written as 
\begin{equation}\label{eq:entropy_stable}
S_e(\alpha, T) = \sum_{l=1}^{Q} \left[ \ln\left(1 + e^{-\varepsilon_l(\alpha)/T}\right) + \frac{\varepsilon_l(\alpha)}{T} f_0 \right].
\end{equation}
This expression avoids numerical instabilities that can arise from computing logarithms of very small numbers at low temperatures.

\section{Thermodynamic Quantities Antisymmetric Butterfly}

The calculation of the electronic specific heat and entropy for the triangular lattice follows similar steps to those in the symmetric butterflies, here including the corresponding $\mu=\mu(\alpha,T)$ at half-filling. The internal energy is given by Eq.\ \ref{eq:internal_energy}. 
The related specific heat is $C_e=\partial U_e/\partial T$.  
We find it convenient to directly compute the specific heat from the numerical derivative of $U_e$, using a small temperature step, $h=0.001t$.

The electronic entropy for the triangular butterfly is correspondingly given by  
Eq.\ \ref{eq:entropy_stable}, with $\varepsilon_l$ replaced by $\varepsilon_l-\mu$, accordingly.

\bibliography{bib}

\end{document}